# Experimental Investigation of Acoustically Forced Helium Jet in Crossflow Using Shadowgraphy and Modal Analysis

Ankur Kumar[1], Sita Ram Sahu[2], Narsing K Jha[3], and Anubhav Sinha[1]*

[1]Department of Mechanical Engineering, Indian Institute of Technology (Banaras Hindu University), Varanasi, India

[2]School of interdisciplinary Research, Indian Institute of Technology, Delhi, India

[3]Department of Applied Mechanics, Indian Institute of Technology, Delhi, India

*Corresponding author email – er.anubhav@gmail.com

## Abstract

*This study presents an experimental investigation of helium jet in crossflow of air, with an objective to understand the influence of acoustic forcing on jet behavior and mixing. Unforced and sinusoidally forced jets are studied. High speed shadowgraphy is used to capture instantaneous jet images. These images are further processed using Proper Orthogonal Decomposition (POD) algorithm to provide insights into the spatiotemporal behavior. Schlieren Imaging Velocimetry (SIV) is also used to understand the jet entrainment and identify regions of low and high velocities. Moreover, an interface tracking method is used to obtain interface location oscillations for near nozzle and far field locations. Energy spectrum is obtained from Fast Fourier Transform (FFT) of these oscillations. The unforced jet is observed for different jet to crossflow velocity ratios (R), and it is observed that the jet penetration increases with R. Unforced cases show a broadband spectrum indicating the absence of any dominant frequency, except for the lowest velocity ratio case. The instability in unforced cases is limited to the shear layer oscillations. For the forced cases, a clear dominant frequency corresponding to the forcing frequency (and occasionally their harmonics) is recorded. It is observed that the impact of forcing function is not the same for all the frequencies, and the jet response is observed to be much more pronounced for a higher frequency as compared*

*to a lower frequency forcing. Two typical cases, highlighting the effect of frequency response are compared in detail using image analysis, interface tracking, POD and SIV.*

**Keywords** – Jet in crossflow, acoustic forcing, Shadowgraphy, POD, SIV, high speed imaging, interface tracking.# 1. Introduction

Jet in crossflow (JIC) is a canonical flow of much significance for a range of engineering applications as wells as natural processes [1-3]. Smoke dispersion from a volcano or a chimney stack in presence of wind can be considered as jet in crossflow. Engineering applications include pollutant dispersion, film cooling, secondary dilution air in gas turbine engines, and thrust vectoring in aerospace applications. JIC is one of the most prominent configurations for fuel injection. In spite of being a simplistic arrangement, the interaction between the jet and the crossflow produces a variety of complex vortex structures, including counter-rotating vortex pairs (CVPs), kidney-shaped vortices, hairpin vortices, and wake vortices, among others [1–3]. Of particular relevance to the present investigation is the shear layer vortices which form at the interface of jet and crossflow and play a significant role in fuel mixing and penetration. For a combustor, the ability of the jet to penetrate and effectively mix with the crossflow will determine the combustion efficiency, flame stability, and pollutant formation. Moreover, JIC is also amenable to flow control strategies like acoustic forcing to pulse the jet, which can enhance penetration and mixing, and can provide a control mechanism independent of operating conditions. However, the forcing parameters need to be optimized and their impact on JIC flow field need to be examined in detail, which will enable to tailor or customize the injection system effectively.

Hydrogen is a potential green fuel, expected to be a dominant fuel in near future, and has gained significant attention from researchers recently. It has higher diffusivity, wider flammability range, higher flame speed, lower ignition energy and higher propensity to flashback than conventional hydrocarbon fuels. However, these properties also raise safety concerns in using hydrogen for experimental investigations. To overcome the safety concerns, while being able to study the underlying physics, helium is often used as a surrogate gas for hydrogen in fundamental studies.

Physical properties of helium match closely with hydrogen, making it an ideal candidate. The use of helium enables detailed exploration of flow instabilities, vortex dynamics, and mixing mechanisms without the added complexity and risk associated with combustion. These similarities in physical properties, particularly in density ratio, justify the use of helium jets to develop insights that can be extended to hydrogen systems, thereby contributing to the broader understanding of fuel–air mixing and flame stabilization in hydrogen-fueled devices.

Owing to its significance in several engineering applications, JIC configuration has been studied extensively in literature. Prior studies have investigated jet trajectory and scaling [4–7], highlighting the importance of the jet-to-crossflow velocity and momentum flux ratios. Jet penetration and mixing characteristics substantially affect combustion efficiency, and these behaviors are often controlled by modulating or forcing the jet, usually through acoustic forcing. Johri et al. [8] explored strategies to enhance mixing for a reacting jet in crossflow. They observed puff formation when the jet was forced and investigated its impact on jet structure and mixing. Eroglu and Breidenthal [9] studied JIC in a water tunnel using the same experimental facility as [8]. They reported significant increase in penetration for a forced jet in JIC, as compared to regular jets. However, using a simplistic arrangement of a speaker controlled with a signal generator has its limitations. The forced jet may not follow the acoustic signal exactly, and will have some distortion. To address this issue, some researchers [10, 11] have developed a dynamic compensator which reduces distortion and shows good match with the forcing function at the jet exit. Shapiro et al. [11] have used this strategy to study mixing and penetration of forced jet in crossflow. Shoji and coworkers conducted extensive experimental studies [12–15] involving acoustically forced jets in crossflow. Using planar laser-induced fluorescence (PLIF) with acetone as a tracer and hot-wire anemometry, they characterized the temporal response of jet-crossflow interactions and explored the lock-in dynamics associated with upstream shear layer instabilities. Poincaré maps were employed to assess lock-in phenomena, with observed frequencies in the kHz range [12]. Further investigations revealed that as the momentum ratio decreased, the shear layer transitioned from convective to global instability [13]. These findings were supported by stereo particle image velocimetry and spectral analysis from hot-wire measurements. The effects of square-wave excitation were explored in [14], where enhanced mixing and penetration were observed under both convectively and globally unstable conditions, particularly at high velocity ratios. They have quantified trends in degree of jet penetration and spread. Stroke length is considered to be an important parameter in their study. They have varied the momentum ratio and stroke number, but the forcing frequency was held constant at 100 Hz. In the present work, we will explore the effect of frequency in detail. Harris et al. [15, 16] employed a distinct double-pulse waveform for jet forcing, leading to the formation of multiple vortex rings with diverse interaction modes—avoidance, interaction, or collision—depending on the forcing characteristics, ultimately influencing mixing dynamics.

Numerical approaches have also contributed significantly to understanding JIC dynamics. Mahesh and co-workers [17,18] performed direct numerical simulations (DNS) and applied Dynamic Mode Decomposition (DMD) to analyze unsteady jet behavior and shear layer instabilities at low velocity ratios (R = 2, 4). Their work highlights the importance of the jet exit velocity profile on overall stability. Getsinger et al. [19] introduced variations in jet density using helium–nitrogen mixtures and demonstrated that lower velocity and density ratios lead to a transition from convective to global instability, as characterized by power spectra from hot-wire measurements. Numerical investigations by Balaji et al. [20] for helium and argon jets further illustrated the influence of density on jet penetration, with empirical correlations proposed for trajectory prediction.

Theoretical frameworks also play a crucial role in understanding shear layer instability. Huerre and Monkewitz [21] laid the foundational criteria for transition between absolute and convective instabilities in shear layers. Experimental and theoretical work by Strykowski and Niccum [22] identified a critical velocity ratio (R = 1.32) marking the transition to absolute instability. Alves et al. [23] conducted linear stability analysis of inviscid jets in crossflow and identified helical mode instabilities triggered by crossflow interaction. Their subsequent work with Kelly [24] extended this framework to include viscous effects, resulting in improved predictions of instability growth rates and wavelengths consistent with experimental observations. Bagheri et al. [25] performed three-dimensional simulations and linear stability analyses for JIC flow fields, revealing self-sustained oscillations at a velocity ratio of 3. Sayadi and Schmid [26] conducted numerical investigations of reacting and non-reacting JIC systems, identifying a pronounced difference in frequency response, with reacting jets responding to lower frequencies. However, these simulations assumed initially laminar inflow conditions, warranting further validation through controlled experiments.

Nair et al. [27, 28] have conducted experimental investigations on reacting jets in crossflow using high-speed laser diagnostics. Their study explores the effects of varying density ratio, momentum ratio, and heat release rate by blending hydrogen with helium and nitrogen. The results indicate that heat release plays a crucial role in jet stability, primarily by suppressing the growth of shear layer vortices. Steinberg et al. [29] examined heated hydrogen jet flames under preheated conditions, revealing two distinct flame stabilization modes in a crossflow configuration: a stable lee-stabilized flame and a dynamically lifted flame. These modes are governed by recirculation zones and the local strain-rate field, with the flame position showing strong sensitivity to the fluid mechanical strain rate. A direct numerical simulation (DNS) study by Kolla et al. [30] further investigated the influence of jet injection angle on flame structure and stability, including blowout limits. While several studies have addressed reacting hydrogen jets in crossflow, a comprehensive understanding of the complex combustion dynamics in this configuration is still lacking. Among the key challenges is flame flashback, which poses significant safety concerns and remains a major barrier to the broader deployment of hydrogen-based combustion systems [31–33]. Previous research has typically identified the natural frequencies of unforced jets and applied these in forcing studies. However, these natural frequencies often exceed the practical limits of acoustic excitation. Consequently, a specialized feedback control system is developed to enable jet forcing at elevated frequencies. In many practical systems—such as gas turbines—instabilities occur at relatively low frequencies (typically below 500 Hz), driven by factors including geometry and flame-flow interactions [34]. Thus, understanding jet behavior under low-frequency excitation is critical. This study explores whether jets can be effectively excited at frequencies distinct from their natural frequencies, the minimum amplitude of forcing required, and whether the jet exhibits preferential response to specific frequency ranges. The experimental study attempts to address these key research questions.

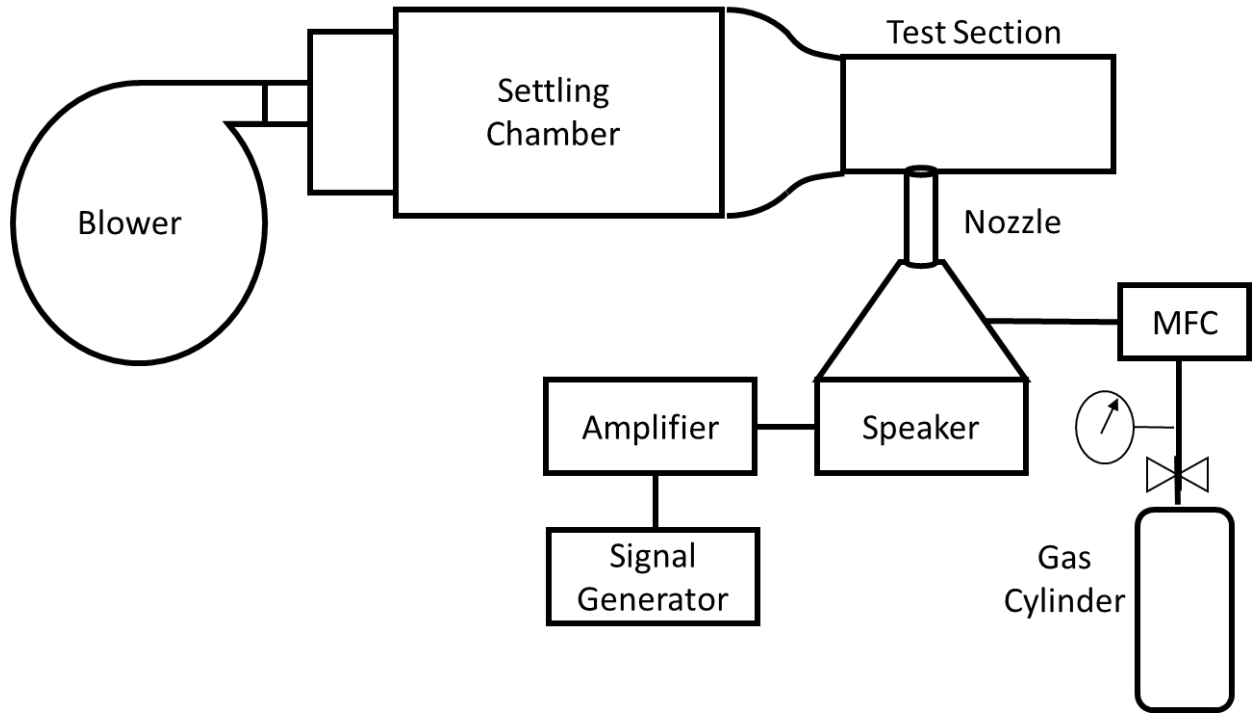

(a) Experimental set-up showing air and gas (Helium) flow pathways

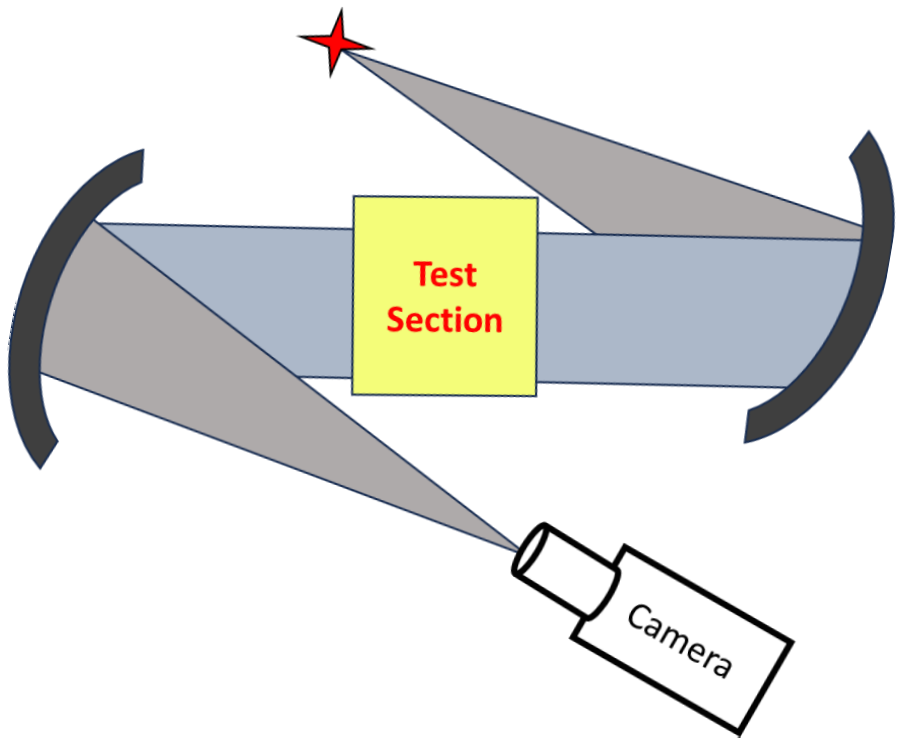

(b) Imaging Technique

Figure 1. Experimental facility and Imaging technique

## 2. Experimental Facility and Image Processing

The experimental setup comprises the air and fuel delivery systems, an acoustic forcing arrangement, and a shadowgraphy-based flow visualization facility. Each of these components is described in detail in the following subsections. Image processing tools are also explained.

### *2.1. Air and fuel flow lines*

A schematic of the experimental facility, including the air and fuel flow lines, is shown in Fig. 1(a). Airflow is generated using a centrifugal blower powered by a 10 HP motor. The motor is driven through a variac, allowing precise control of the blower's rotational speed and, consequently, the airflow rate. Air velocity at the test section is monitored using Pitot tube measurements. The blower outlet is connected to a settling chamber, which helps condition the flow before it enters a converging section. This converging section is designed with a fourth-order polynomial profile to promote a uniform and stable flow. It is directly connected to the test section, which has dimensions of 105 mm in height, 50 mm in width, and 300 mm in length. Quartz windows are mounted on the side walls to facilitate optical access for imaging. A fuel injector is flush-mounted at the center of the bottom wall of the test section. Helium is supplied from a high-pressure cylinder through stainless steel pipelines, equipped with control components such as pressure regulators, safety valves, and mass flow controller (Make- Alicat, model- cori flow) to ensure precise delivery.

### *2.2. Acoustic forcing*

Acoustic forcing is implemented using a function generator (Make-Siglent Model-SDG-1032X), which supplies the desired frequency signal to a voltage amplifier. The amplified signal drives a loudspeaker (Make- JXL, model-1090) housed within a sealed conical chamber, as shown in Fig. 1(a). The chamber is designed to isolate the acoustic source and includes a side-wall fuel inlet, with its top end connected directly to the fuel injector. The injector has a contoured internal surface and an exit diameter of 4 mm. The frequency of acoustic excitation is set via the function generator, while the amplitude is adjusted through the amplifier. The amplitude of oscillations ($\alpha$) at the nozzle exit is characterized using a hot-wire anemometer under no-crossflow conditions.

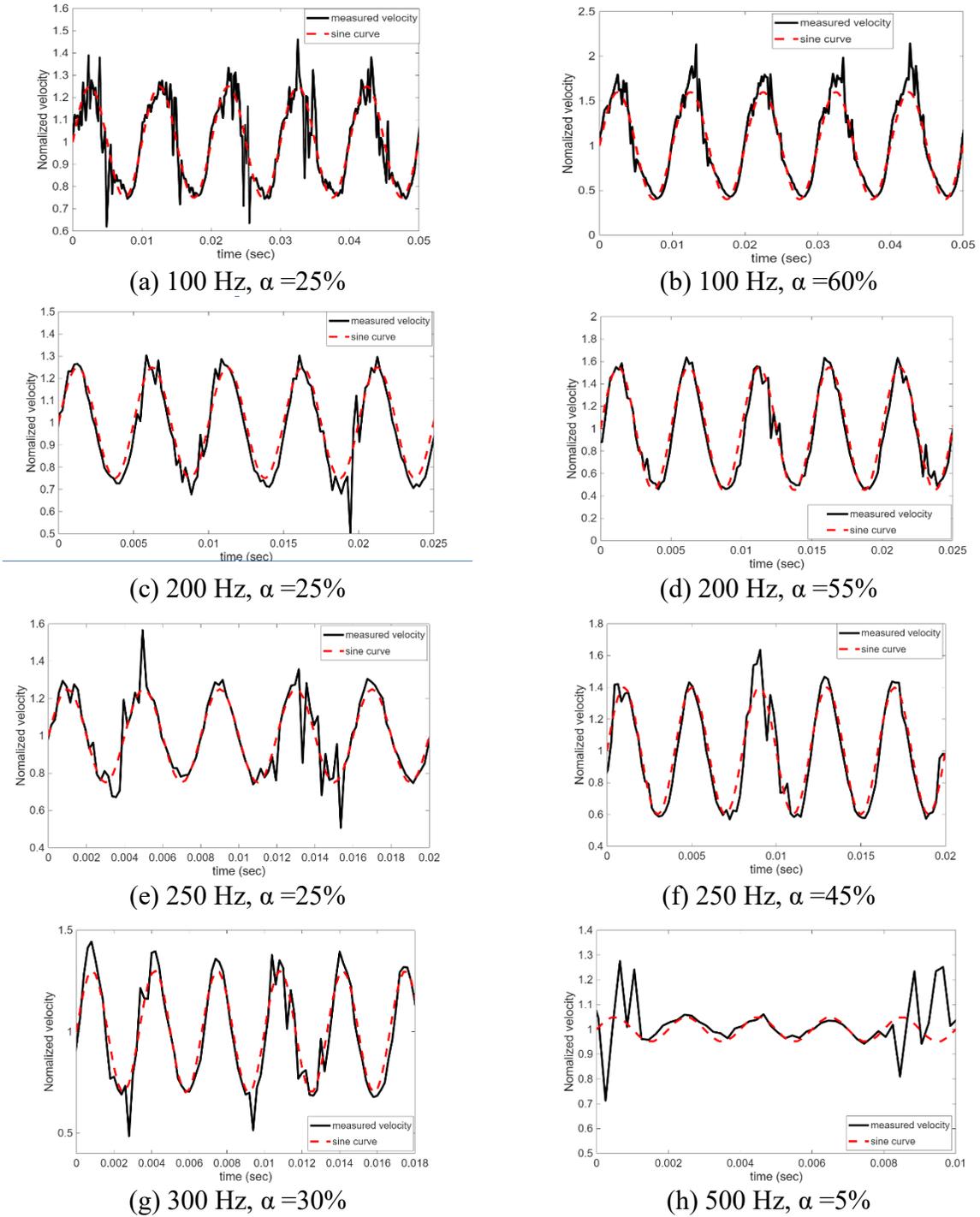

**Figure 2. Normalized velocity overlapped with forcing function for various forcing conditions.**

Corresponding velocity measurements are shown in Fig. 2. As observed, the jet exit velocity shows a good match with the forcing signal characteristics. The forcing function is in the form:

$$U = U_0(1 + \alpha \sin(\omega\, t)) \tag{1}$$

where $U_0$ is the average velocity (without forcing), $\alpha$ is the amplitude of forcing, $\omega$ is the angular frequency, and $t$ is time.

*2.3 Shadowgraphy*

The schematic of the shadowgraphy-based imaging setup is shown in Fig. 1(b). The optical system comprises two 6-inch diameter spherical mirrors (make- Edmund Optics) placed around the test section, as illustrated in Fig. 1. A high-intensity LED light source, enclosed in a box with a pinhole aperture to serve as a point source, is positioned at the focal point of the first mirror. Upon reflection, the light rays become collimated and pass through the test section. These parallel rays are then reflected by the second spherical mirror and focused onto a high-speed camera (Photron mini AX-100) placed at the focal length of the second mirror. The presence of density gradients in the flow field induces local variations in refractive index, causing deflection of the parallel light rays. These deflections are captured by the camera, enabling visualization of the helium jet and its dispersion. All images in this study were acquired at a frame rate of 10,000 frames per second.

*2.4. Proper Orthogonal Decomposition (POD)*

Proper Orthogonal Decomposition (POD) is an advanced data analysis technique used to extract coherent structures from complex flow fields. It identifies the most energetic flow features and their associated frequencies, making it a powerful tool in fluid mechanics research. POD mode shapes represent the dominant dynamic structures present in the flow, while accompanying Power Spectral Density (PSD) plots help identify the characteristic frequencies associated with these modes. By compressing large volumes of experimental or computational data into a reduced set of modes ranked by energy content, POD enables insightful interpretation of unsteady flow physics. For a comprehensive understanding of the methodology, advantages, and wide-ranging applications of POD, readers are referred to foundational review articles [35, 36]. Some applications of POD based analysis of jet in crossflow configuration has been shown in [37, 38].

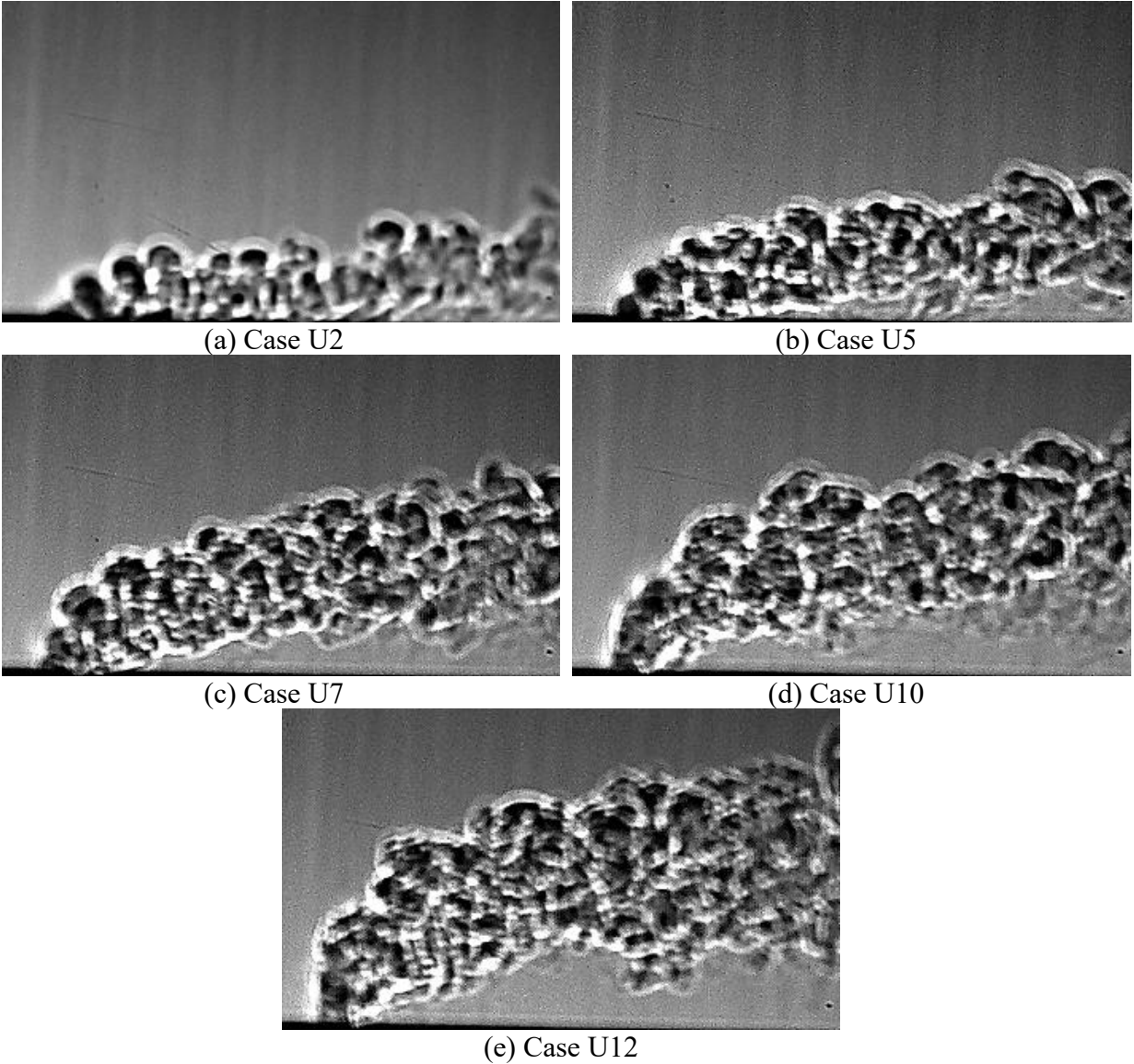

(a) Case U2  (b) Case U5
(c) Case U7  (d) Case U10
(e) Case U12

**Figure 3. Instantaneous images of unforced cases at different velocity ratios.**

| Case | Velocity ratio (R) | Momentum flux ratio (q) | Jet Reynolds number (Re) |
|---|---|---|---|
| U2 | 2.4 | 0.9 | 728.2 |
| U5 | 4.9 | 3.5 | 1456.3 |
| U7 | 7.3 | 8.0 | 2184.5 |
| U10 | 9.8 | 14.2 | 2912.7 |
| U12 | 12.2 | 22.1 | 3640.8 |

**Table 1. Unforced cases and corresponding parameters**

| Case    | Forcing Frequency (Hz) | Amplitude $\alpha$ (%) |
|---------|------------------------|------------------------|
| F100-25 | 100                    | 25                     |
| F100-60 | 100                    | 60                     |
| F200-25 | 200                    | 25                     |
| F200-55 | 200                    | 55                     |
| F250-25 | 250                    | 25                     |
| F250-40 | 250                    | 40                     |
| F300-30 | 300                    | 30                     |
| F500-5  | 500                    | 5                      |

**Table 2. Forced cases and their forcing parameters**

## *2.5. Schlieren Image Velocimetry (SIV)*

Schlieren Image Velocimetry (SIV) is an advanced image-based flow diagnostic technique derived from Particle Image Velocimetry (PIV) algorithms [39-41]. Unlike PIV, which relies on the presence of tracer particles in the flow, SIV operates on Schlieren or shadowgraphy images that visualize refractive index gradients caused by density variations. The method involves cross-correlating regions of interest in consecutive images to estimate the displacement of coherent flow structures, typically along the direction of dominant gradients such as shear layers, vortices, or shock fronts. This allows for the estimation of planar velocity fields in flows where traditional seeding is not practically possible or undesirable. SIV is particularly well-suited for compressible and reacting flows, where density gradients enable optical methods like Shadowgraphy to capture flow features. In the present study, SIV is employed to process the Schadowgraphy images obtained for helium jets in crossflow. Opensource code PIVlab is used [41] for this study.

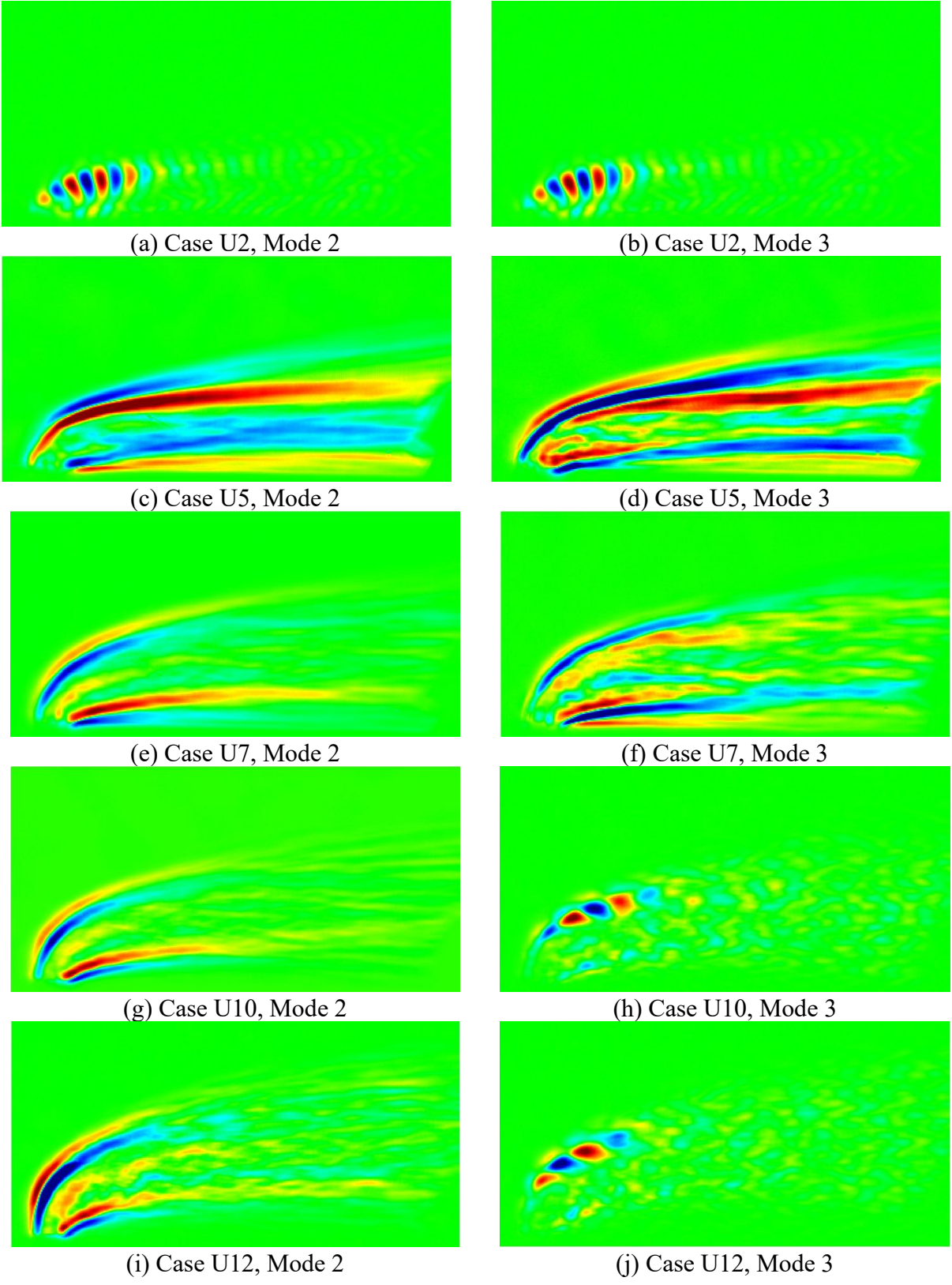

**Figure 4. POD modes for unforced cases**

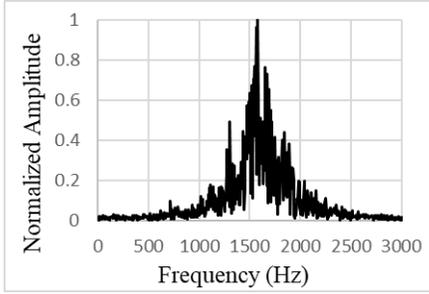
(a) Case U2, Mode 2

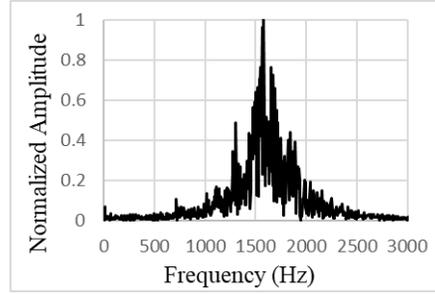
(b) Case U2, Mode 3

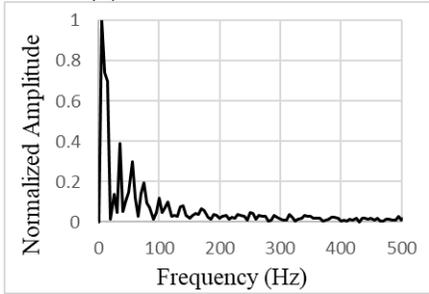
(c) Case U5, Mode 2

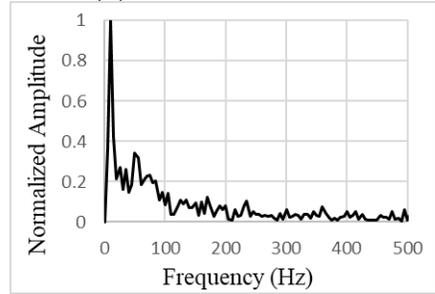
(d) Case U5, Mode 3

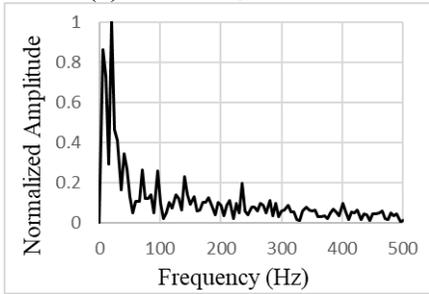
(e) Case U7, Mode 2

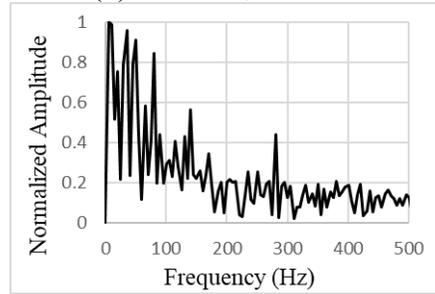
(f) Case U7, Mode 3

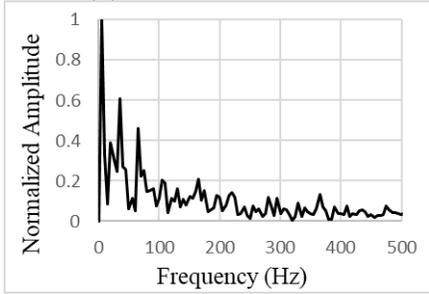
(g) Case U10, Mode 2

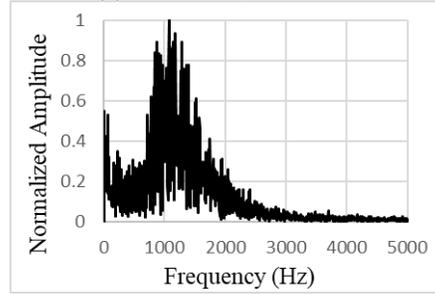
(h) Case U10, Mode 3

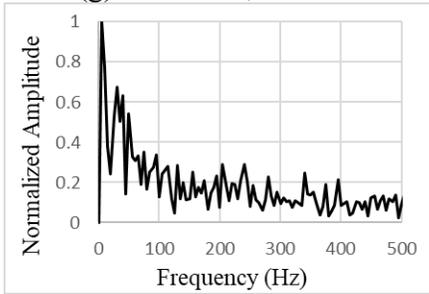
(i) Case U12, Mode 2

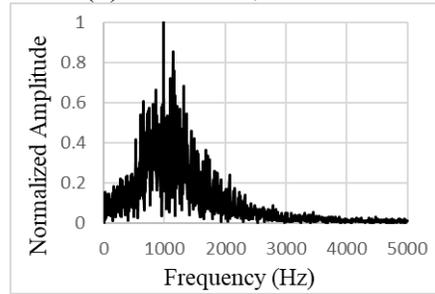
(j) Case U12, Mode 3

**Figure 5. PSD plots for unforced cases**

# Results

## *3.1. Unforced cases*

The crossflow air velocity is held constant at 8.2 m/s for all the experiments reported in this study. The helium jet velocity is varied to obtain different velocity ratios. Unforced cases along with the corresponding velocity ratios are listed in Table 1. Forced case details are given in Table 2. Velocity ratio ($R$) can be defined as the ratio of jet velocity (averaged) and crossflow velocity.

$$R = \frac{U_{jet}}{U_{air}} \tag{1}$$

where $U_{jet}$ is the jet velocity and $U_{air}$ is the crossflow air velocity. Another important parameter is the momentum flux ratio ($q$) which can be expressed as:

$$q = \frac{\rho_{jet} U_{jet}^2}{\rho_{air} U_{air}^2} \tag{2}$$

where $\rho_{jet}$ is the jet density and $\rho_{air}$ is the crossflow air density. Jet Reynolds number is based on jet diameter and listed in Table 1. Helium density is taken as 0.1784 kg/m³, and viscosity as $1.96 \times 10^{-5}$ N. s/ m² for these calculations. Air density is assumed to be 1.2 kg/m³.

Instantaneous images of the unforced helium jet in crossflow for various cases are shown in Fig. 3. It is evident that the jet penetration increases with increase in velocity ratio ($R$) which is intuitive. For the lowest velocity ratio case (U2), the jet crawls along the bottom plate and the lee-ward boundary cannot be defined. Whereas, for the higher R cases, the lee-ward boundary is clearly defined, as it raises higher than the bottom plate. The wind-ward trajectory depicts large scale structures which form due to interaction between the jet and the crossflow. These structures are clearly visible for the U10 and U12 cases. Overall, the jets appear to be turbulent and a large number of vortical structures are observed in the jet flow field. To understand the spatio-temporal behavior of these jets, POD analysis is carried out. The POD mode shapes are presented in Fig. 4 and corresponding PSD plots are shown in Fig. 5. Only the most prominent modes, Mode 2 and Mode 3 are presented. Mode 1 which captures the average behavior is not included as the focus of the present work is on dynamic behavior. U2 appears very different from other cases. The POD mode shape appears like a train of structures moving in the direction of crossflow. These structures

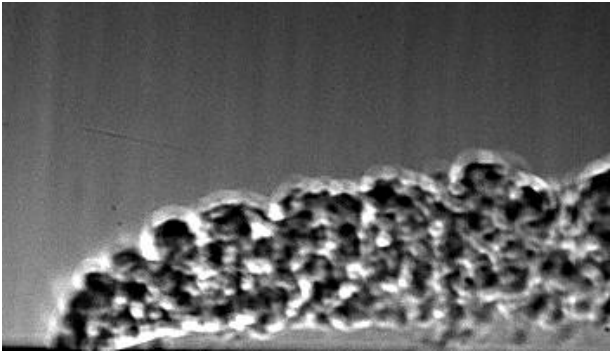
(a) Case F100-25: 100 Hz, α=25%

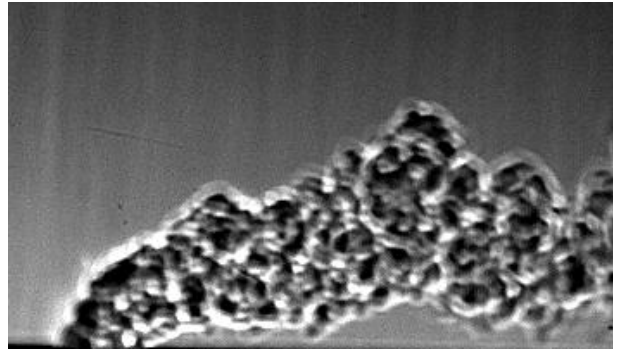
(b) Case F100-60: 100 Hz, α=60%

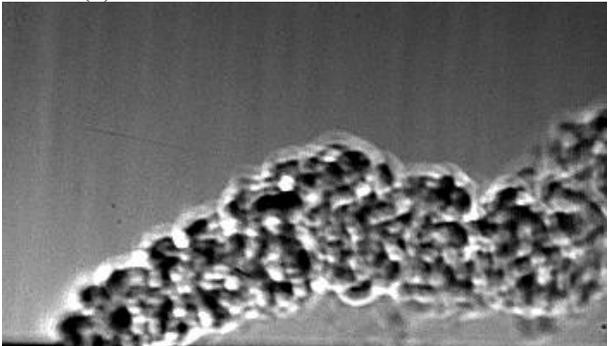
(c) Case F200-25: 200 Hz, α=25%

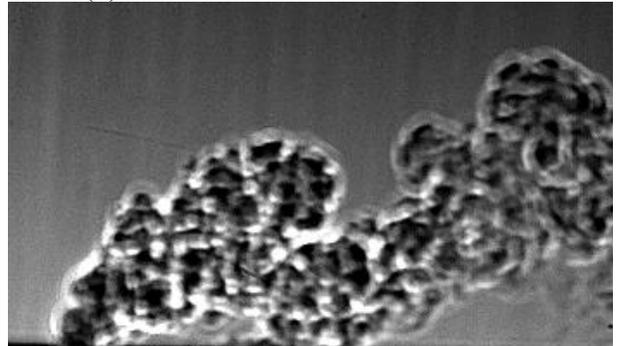
(d) Case F200-55: 200 Hz, α=55%

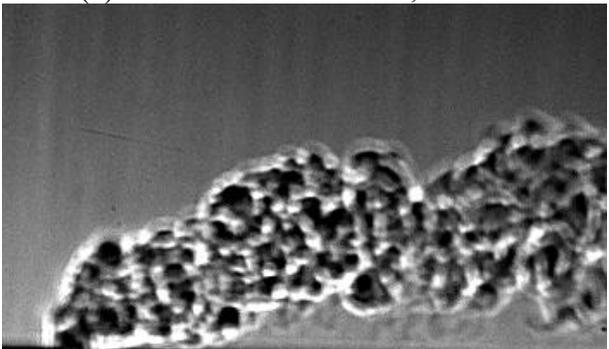
(e) Case F250-25: 250 Hz, α=25%

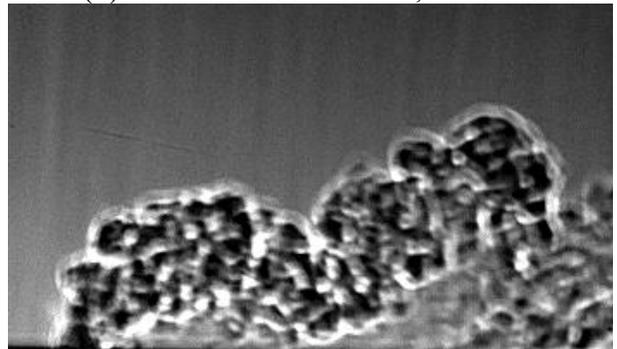
(f) Case F250-40: 250 Hz, α=40%

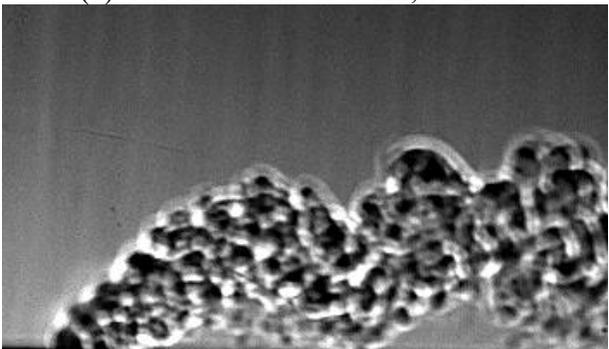
(g) Case F300-30: 300 Hz, α=30%

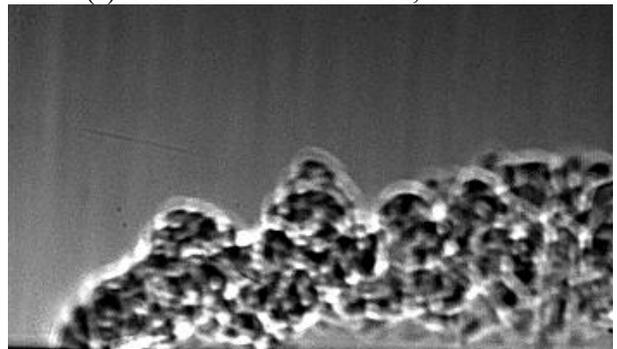
(h) Case F500-5: 500 Hz, α=5%

**Figure 6. Instantaneous images of forced cases**

are also formed at high frequency as evident in their PSD plots. In contrast, the mode 2 plots of other cases display oscillations in the shear layer region of both wind-ward and lee-ward boundaries. This feature is characterized by very low frequencies as shown in the PSD plots. The jet to crossflow interaction is visible in a vigorous form in the wind-ward shear layer, where the shear layer gets perturbed. However, at the lee-ward side, the effect of counter rotating vortex pair (CVP) is becoming apparent which is revealed in the POD mode shapes. This has been confirmed with the high-speed images.

Mode 3 of U2 case shows the same structure as observed in mode 2, with same frequency plot, showing the dominance of this feature. Mode 3 for U5 and U7 cases show perturbation in the shear layer. The frequency response is noisy with several peaks. This highlights the complex nature of shear instability in these cases with interaction of structures of various characteristic length and time scales. The lee-ward shear layer again shows signs of CVP induced motion. The higher R cases show the perturbation of the wind-ward shear layer at high frequency (around 1000 Hz). As these are the highest R cases of this study, the interaction between the jet and crossflow air is expected to be the strongest in these cases. This interaction is so strong that it has shadowed the CVP induced motion in the lee-ward boundary, and only a train like alternating structure is visible in the third mode.

*3.2. Forced cases*

Further, the R is held constant at 7.3 and these cases are forced with various frequencies at different amplitudes as discussed in the previous section. The impact of forcing function is measured using a hot wire probe mounted at the nozzle exit without crossflow. Measured velocities are normalized and superimposed with the sinusoidal function and shown in Fig. 2. A few cases have some noise, but overall, the sine function is observed. The impact of forcing is minimal for the 500 Hz case and a very weak signal is observed. For higher frequencies, the speaker actuation is too fast to the jet to respond and higher frequencies were not tested. For this paper, forcing up to 500 Hz is studied. As evident from these plots that the 100 Hz and 200 Hz case show similar jet behavior without crossflow. However, these jets show significantly different behavior when subjected to a crossflow. Figure 6 shows typical instantaneous images for forced cases. All these cases are for the same velocity ratio ($R$=7.3). Note than the high amplitude forcing is more effective in jet deformation and mixing as compared to the low amplitude case at the same frequency. It is also observed that for the low frequency case (100 Hz), the jet distortion is not as pronounced as for the higher frequency cases. Even for 5% α case at 500 Hz (Fig. 6(h)) shows better response than the 25% α case for 100 Hz. Cases F100-60 and F200-55 offer an interesting comparison and will be discussed in following section. Figure 7 shows the sequential instantaneous image for cases F100-60 and F200-55. The arrow marks a typical flow structure being tracked as it moves along the crossflow direction. It is observed that the forcing significantly impacts the jet structure for the 200 Hz case (F200-55), while the 100 Hz case (F100-60) appears to receive a moderate response from the jet-crossflow interaction. This is surprising since the perturbation (α) is slightly higher for 100 Hz, and the difference in their forcing frequency is not significant. This shows the JIC configuration weakens the response for lower frequency (100 Hz) while strengthens the response for higher frequencies (≥200 Hz). Some other cases are studied for higher frequencies (250 Hz, 300 Hz, cf. Table 2). However, their response is very similar to the 200 Hz case and hence only 200 Hz case is discussed further for brevity. To further probe these two cases, instantaneous images are processed using SIV algorithm to obtain the horizontal velocity component (u). These processed images (shown in Fig. 8) aid to track the high-speed and low-speed region within the jet moving downstream. The difference in jet structures is also clearly highlighted. Jet exit will have a high value in the v velocity plots, but in u velocity plots, a low-velocity region is visible for jet exit. High velocity regions are observed near the wind-ward trajectory of the jet, near the shear layer. They gradually convect along the crossflow.

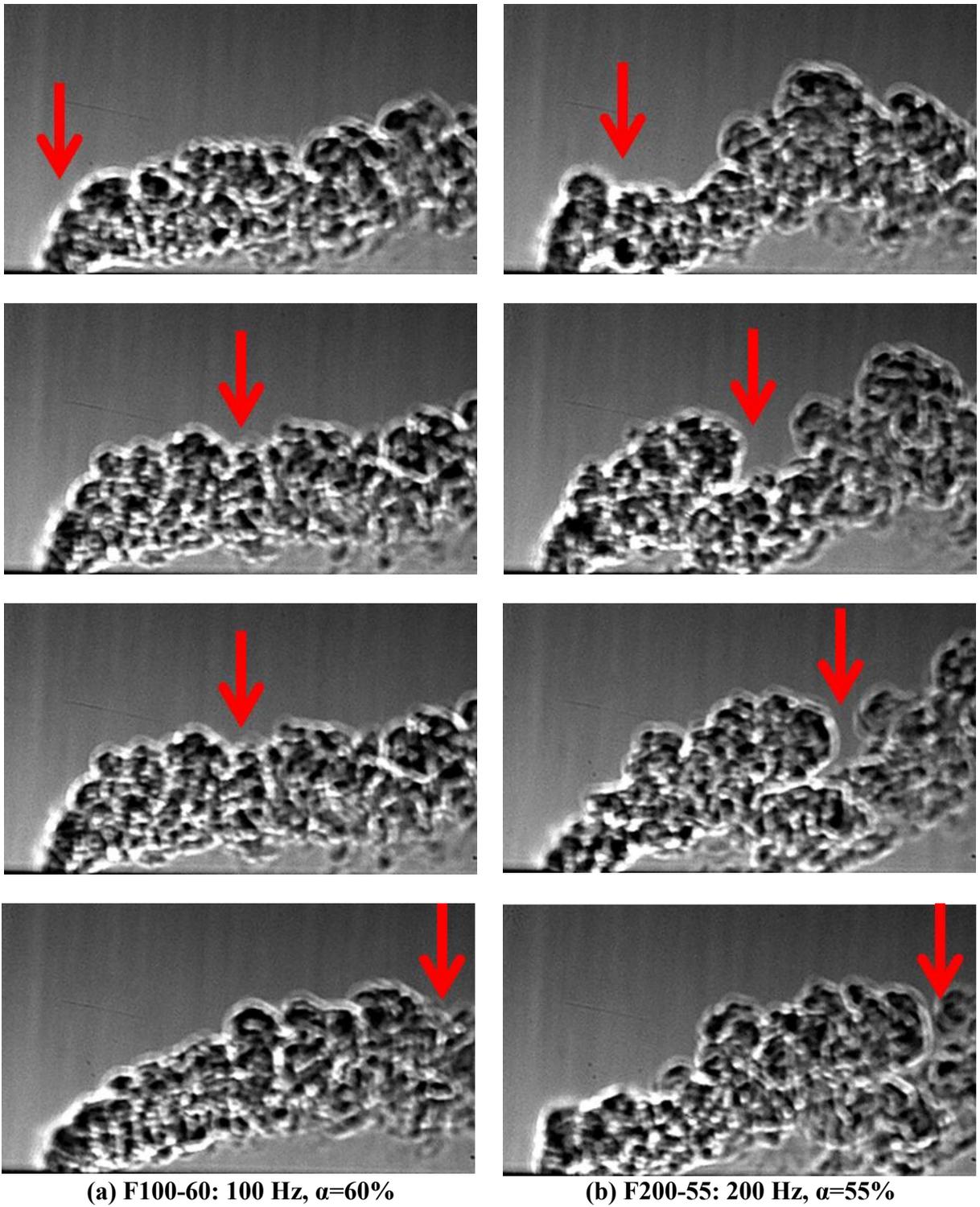

**(a) F100-60: 100 Hz, α=60%**  **(b) F200-55: 200 Hz, α=55%**

**Figure 7. Sequential instantaneous images compared for cases F100-60 (left column) and F200-55 (right column)**

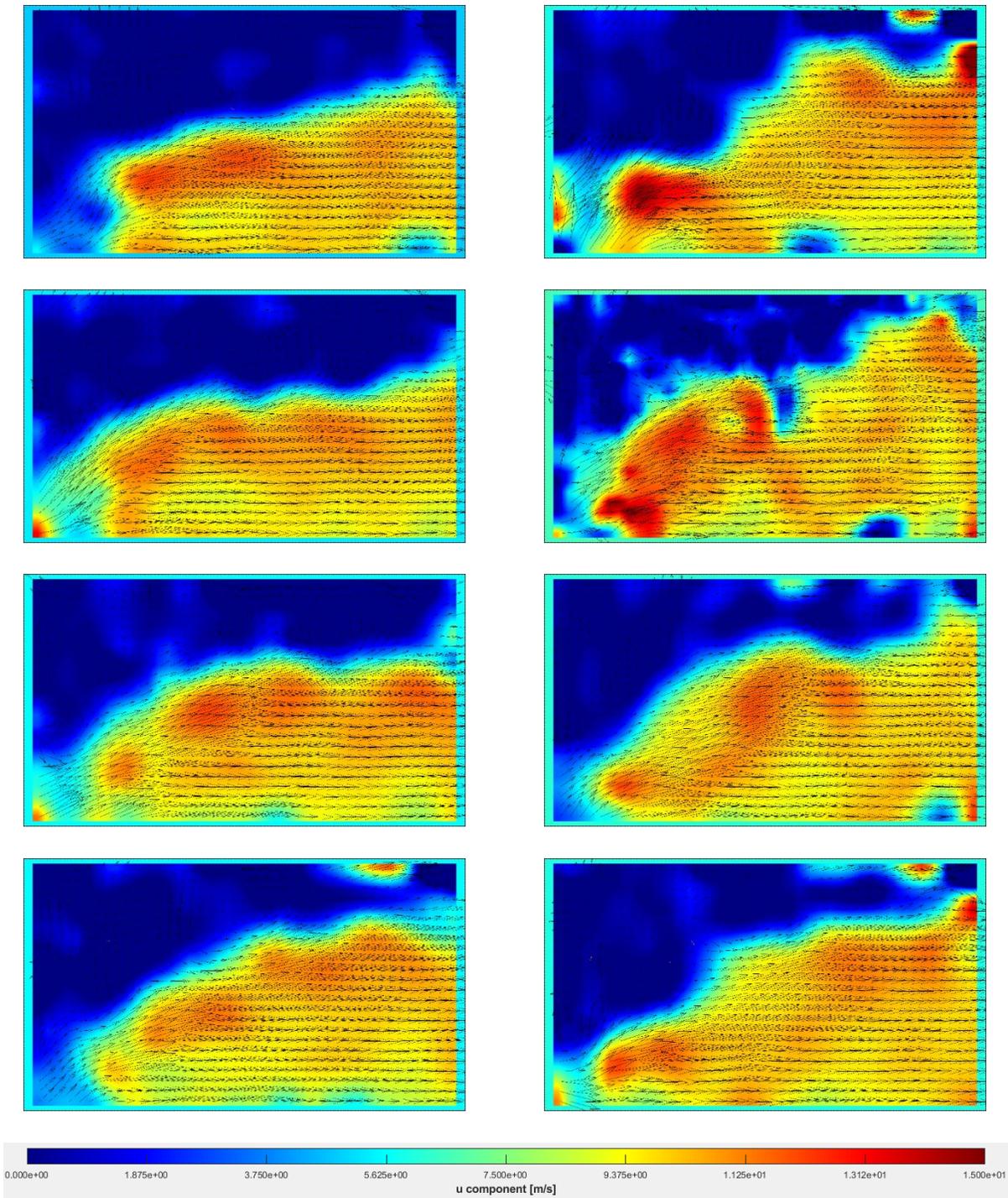

**(a) F100-60: 100 Hz, α=60%**         **(b) F200-55: 200 Hz, α=55%**

**Figure 8. SIV Processed images for cases F100-60 and F200-55.**

To gain further insights into these cases, POD analysis is conducted. POD mode shapes are shown in Fig. 9, and PSD plots are presented in Fig. 10. For PSD plots, the observed frequency is normalized by the forcing frequency. These figures cover cases for both the values of α for 100 Hz and 200 Hz (cf. Table 2). 500 Hz case is also included in these set of images. All the cases show a clearly pronounced peak for the forcing frequency. For 100 Hz cases, the smaller α case (F100-25) shows a clear peak for forcing frequency for PSD of both modes. However, the signal is very noisy, especially for mode 3. This case also shows a significant peak near the zero frequency for both the modes, which is not shown by other cases. For higher α case (F100-60), the peaks are more clear. This shows the effect of increasing α. However, they are noisier as compared to other, higher frequency cases. Interestingly, the third mode of this case shows a peak not at 100 Hz, but at 200 Hz, which further demonstrates that crossflow prefers higher frequencies, particularly 200Hz. For all other modes and all other cases, there is always a dominant, well-defined peak at the frequency of forcing function, and much less noise as compared to 100 Hz cases.

POD mode shapes are shown in Fig. 9. The 100 Hz case (Fig. 9(a)-9(d)) show perturbation at the shear layers. The instantaneous images of these cases show that for 100 Hz case, the jet boundary is not much distorted and remains close to the average shape (cf. fig. 7, 8). Mode 2 and 3 show long streaks in shear layer. Corresponding images of 200 Hz cases (Fig. 9(e)-9(h)) show large scale distortions of the jet boundary which is also observed in instantaneous images. The boundary forms a wavy pattern which is represented by the structure shown in mode 2 and 3. The same features are observed in both modes of both 200 Hz cases. As the is increased, the jet distortion increases which is manifested as the more pronounced mode shape for the F200-55 case as compared to F200-25 case (Fig. 9(e)-9(h)). It is also interesting that the third mode of 100 Hz case (F100-60) looks similar to 200 Hz mode shapes, with the alternating wavy pattern being visible. This is the same mode which shows PSD peak at 200 Hz. This observation further supports the fact that the crossflow is suppressing the lower frequency (100 Hz) and promoting the higher frequency (200 Hz). This will further be validated by analyzing the results for higher frequency case F500-5. Although the forcing frequency is very weak, the jet responds well to this forcing. Please note that the low α for 500 Hz case is not intentional. The challenge is that the acoustic system does not responds equally to all frequencies, and only a low α can be obtained for 500 Hz. In spite of that, the PSD peaks for 500 Hz case are defined very sharp, similar to 200 Hz cases; and both the modes show the dominance of the forcing frequency (500 Hz).

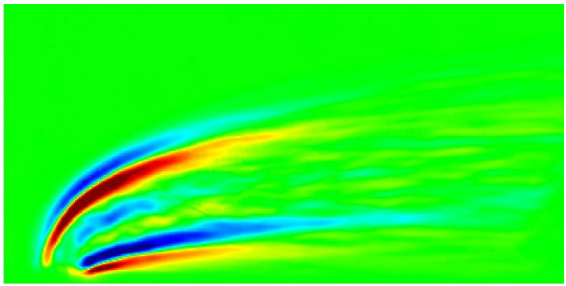
(a) Case F100-25, Mode 2

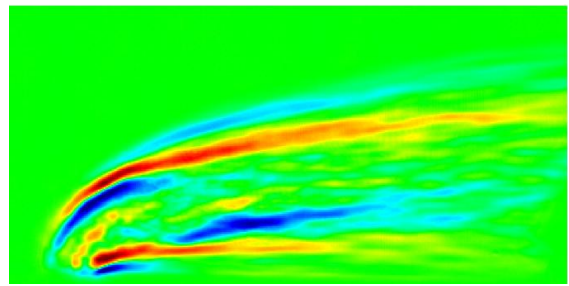
(b) Case F100-25, Mode 3

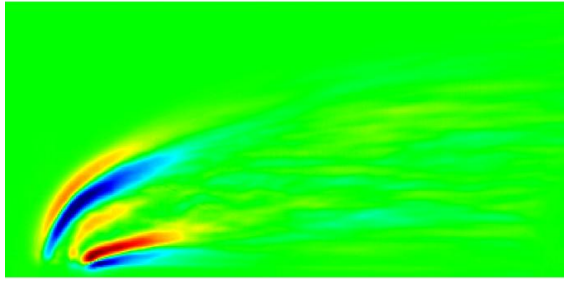
(c) Case F100-60, Mode 2

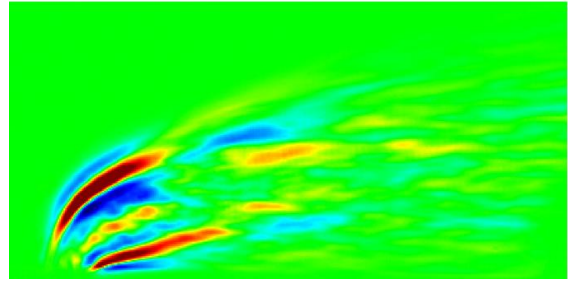
(d) Case F100-60, Mode 3

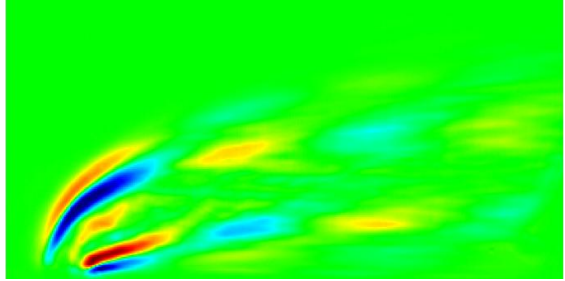
(e) Case F200-25, Mode 2

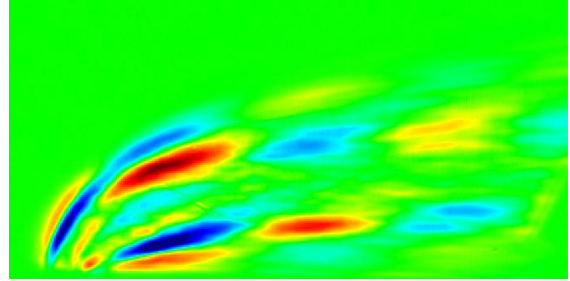
(f) Case F200-25, Mode 3

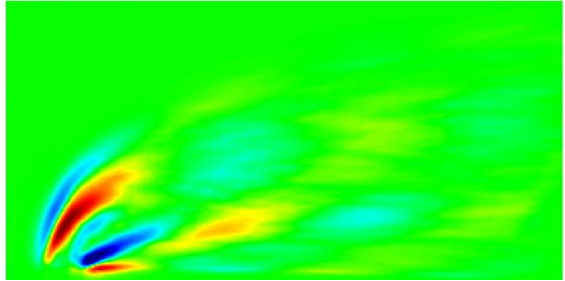
(g) Case F200-55, Mode 2

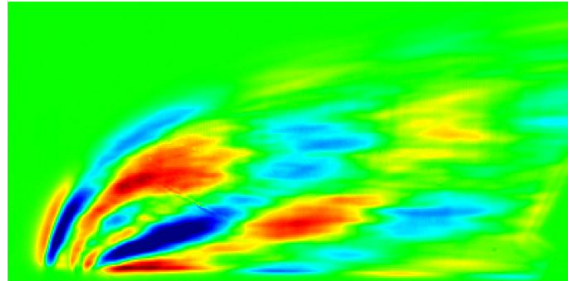
(h) Case F200-55, Mode 3

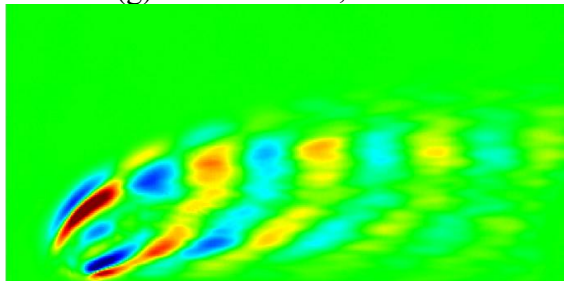
(i) Case F500-5, Mode 2

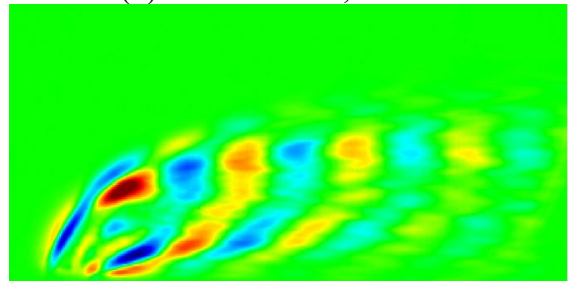
(j) Case F500-5, Mode 3

**Figure 9. POD mode shapes for forced cases.**

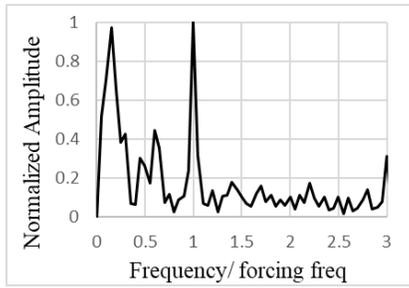
(a) Case F100-25, Mode 2

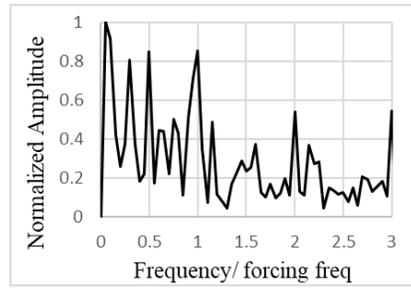
(b) Case F100-25, Mode 3

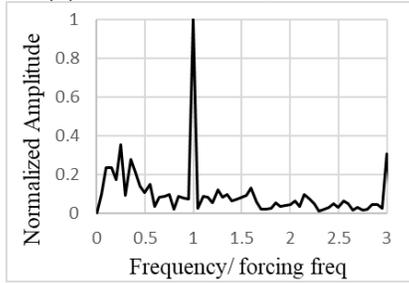
(c) Case F100-60, Mode 2

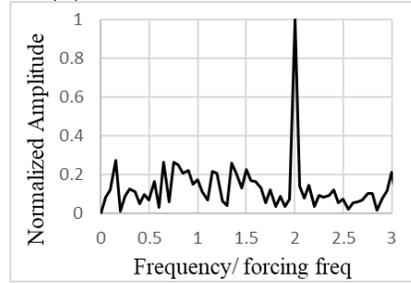
(d) Case F100-60, Mode 3

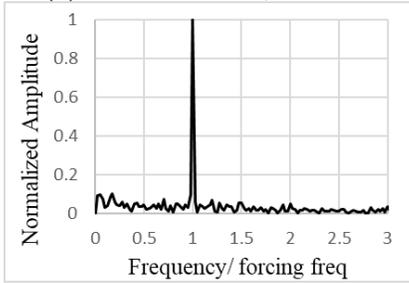
(e) Case F200-25, Mode 2

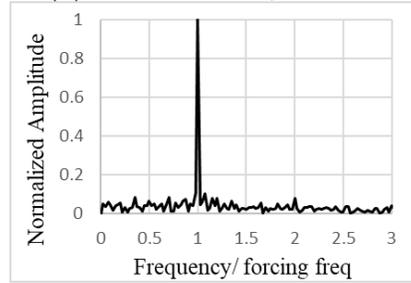
(f) Case F200-25, Mode 3

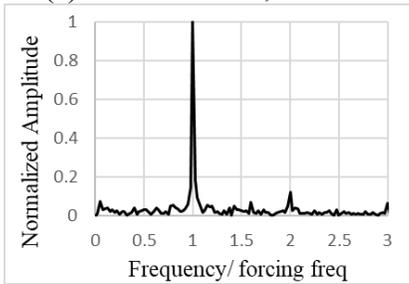
(g) Case F200-55, Mode 2

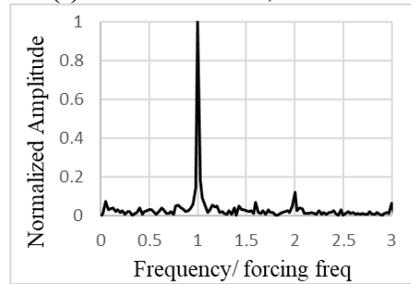
(h) Case F200-55, Mode 3

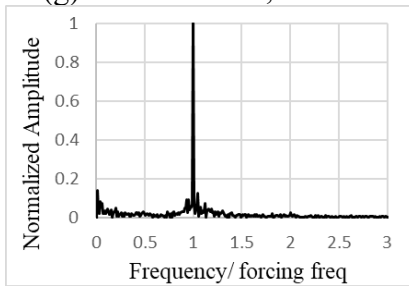
(i) Case F500-5, Mode 2

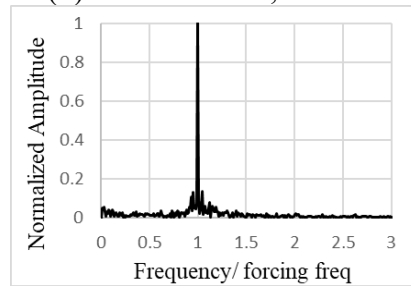
(j) Case F500-5, Mode 3

**Figure 10. PSD plots for forced cases**

For this case, there is no large-scale distortion of the jet. The periodically puffed jet interacts mildly with the crossflow, but the response is strong enough to give a sharp peak at 500 Hz and produce train like structures in POD mode shapes. These structures originate because of forcing only and in spite of very low α, their impact is evident in the POD mode shapes.

### 3.3. Interface Analysis

Jet interface at the wind ward boundary is studied using a contrast-based image processing method using MATLAB. Interface is tracked at two points, which are at a downstream distance $5d_j$ and $15d_j$ from the nozzle exit in the crossflow direction, where $d_j$ is the nozzle exit diameter. The interface point oscillations with time are studied, and a FFT of this oscillation provides the Energy spectra which is discussed in this sub-section. Figure 11 shows the development of energy spectra along the stream of jet i.e. at $5d_j$ and $15$ $d_j$. Energy spectra for F100-60 and F200-55 cases are presented. At $5d_j$ which is closer to the jet exit, the flow is relatively organized due to forcing being dominant. The energy spectra in frequency domain shows clear, sharp peaks at the forcing frequencies. This implies that vortices and interface waves form at regular, periodic intervals, corresponding to the forcing function frequency. It is important to note that while F200-55 case shows peak for the forcing frequency (200 Hz), the F100-60 case also shows a smaller peak for higher frequencies, in addition to its forcing frequency (100 Hz). This observation confirms that the interfacial structures are not only periodic but also contain higher-order modes generated by nonlinear interactions near as well as far downstream of jet. Further, for both the cases, the energy around these dominating frequencies is highly concentrated, with less energy in the higher-frequency range. This suggests that larger coherent structures are dominant, and small-scale fluctuations are not of much influence. This complies with the observation of frequencies in the PSD plots. At 15dj, the flow has evolved significantly as we observed downstream. The energy distribution shows a noticeable shift upward in the spectrum, implying that more energy is content across a broader range of frequencies.

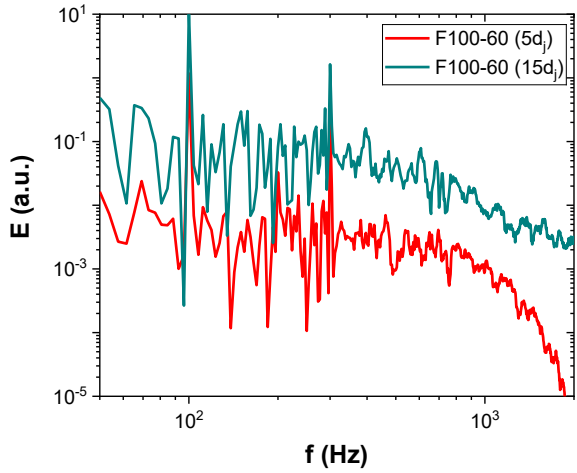
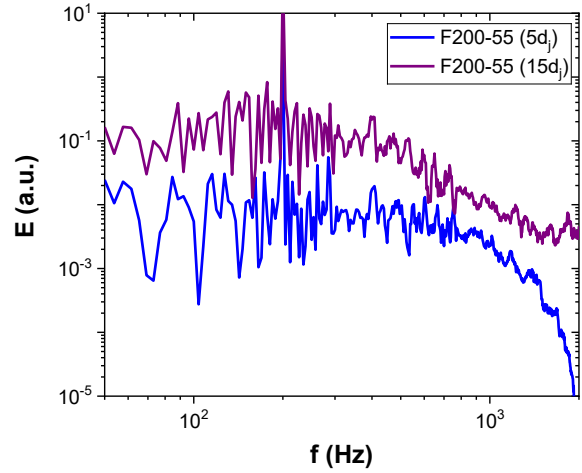

(a) Case F100-60

(b) Case F200-55

**Fig. 11 Energy spectrum in frequency domain for F100-60 and F200-55 cases**

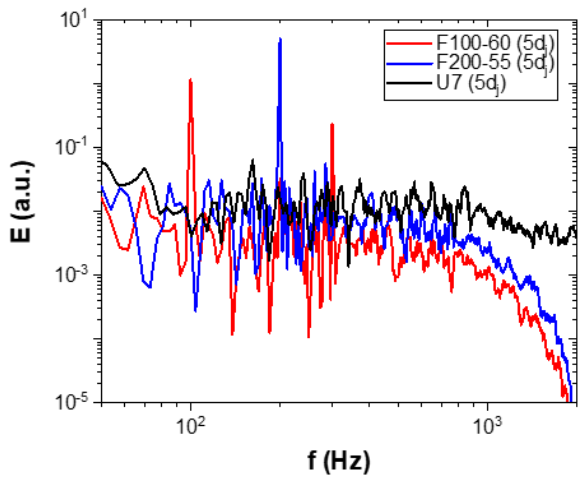
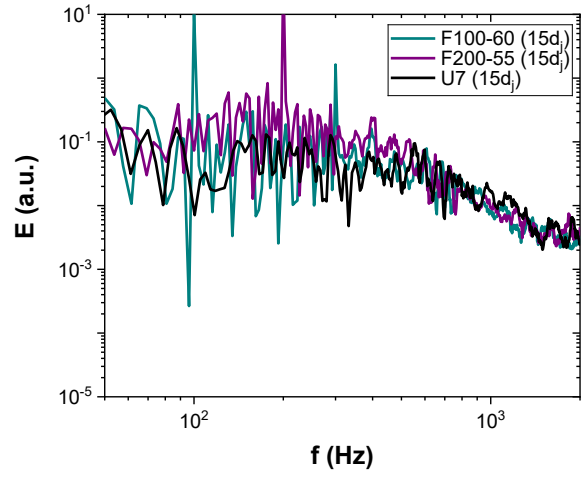

(a) At location 5 $d_j$

(b) At location 15 $d_j$

**Fig. 12 Energy spectrum in frequency domain for F100-60 and F200-55 cases compared with the unforced case U7.**

This shows the development of smaller-scale structures downstream, which gain energy from the breakdown of larger structures representing energy cascading. As turbulence develops, these small-scale eddies extract energy from the mean flow, resulting in enhanced small-scale development. Consequently, the energy spectra for both 5dj and 15dj shows a broader spectral content with logarithmic or algebraic decay at high frequencies, characteristic of turbulent flows and shear layer instability [2, 5, 19]. Energy spectra mark a clear evolution from organized, large-scale dominated flow to more chaotic, small-scale structures, as we go downstream, illustrating classical features of turbulent transition and energy redistribution in jet flows under external forcing.

The unforced case U7, forced cases of F100-60 and F200-55 are compared for their spectra at both the observation locations in Fig. 12. The U7 unforced case, where no external periodic forcing is applied, displays a broad frequency spectrum without a distinct peak. This indicates the absence of any dominant periodic behavior in the interfacial motion. The energy is distributed across a wide range of frequencies, suggesting that vortex shedding is governed by natural instabilities [5]. There are no observable harmonics, and the flow near the jet retains a more chaotic or irregular structure. Small-scale structures begin to form, but without a consistent energy source, they do not dominate the flow dynamics. In contrast, the F100-60 and F200-55 cases introduces a periodic forcing at 100 and 200 Hz. The resulting energy spectra plot in frequency domain shows a strong and sharp peak at the forcing frequency, clearly indicating that the flow has locked into this externally imposed frequency. In addition to the primary peak, harmonics (multiples of 100 and 200 Hz) appear prominently, which confirms that the interfacial structures are not only periodic but also contain higher-order modes generated by nonlinear interactions near as well as far downstream of jet. Most of the energy is concentrated around the forcing frequency and its harmonics, leading to a reduction in energy content at other frequencies. This implies that the flow is being efficiently organized into large-scale coherent structures that dominate the interfacial motion. The logarithmic decay at higher frequencies has been observed which further shows that beyond the main structures, only limited energy is transferred to smaller scales, implying partial suppression of turbulence. The flow behaviour at 5dj and 15dj downstream of the helium jet exhibits significant differences due to the evolution of structures as they travel farther from the jet exit. As discussed, near to the jet the interface dynamics are more directly influenced by the initial forcing or natural instabilities. At this stage, small-scale structures have only begun forming, and the flow retains much of its organized nature. Whereas at far downstream the flow has had more distance and time to evolve. At far downstream energy cascading leads to more active small-scale turbulence which is clear from experimental results. Also, it is observed that at downstream distance the spectra become broader, and sidebands develop, indicating nonlinear interactions and energy transfer to finer scales.

## Conclusions

This experimental study examined the dynamic behavior of a helium jet in crossflow, investigating the effects of acoustic forcing on jet structure and response. Using high-speed shadowgraphy, detailed insights into the spatiotemporal evolution of the jet under both unforced and forced conditions were obtained. The jet is visualized by high speed Shadowgraphy. Jet instantaneous images are examined, and processed using a POD algorithm. POD mode shapes and PSD spectrum are used to understand the spatial modes and their temporal characteristics. Jet penetration is observed to increase with increasing velocity ratio. Unforced jet is found to mostly exhibit features of shear layer instability. The PSD plots mostly exhibit broadband response and only for one case, at the lowest velocity ratio a dominant frequency peak is observed around 1500 Hz. The POD modes highlight the zones of shear layer interaction where the jet interacts with the crossflow. On the other hand, the forced cases show very clear frequency peaks matching the forced frequency. It is observed that the jet response is diminished for 100 Hz, whereas the effect of forcing is quite dominant for 200 Hz and higher frequencies. This could be attributed to the tendency of Kelvin-Helmholtz instability to be more effective for higher frequencies. However, the jet response to acoustic forcing diminishes at higher frequencies and frequency up to 500 Hz can only be examined. Further, jet interface is tracked at two downstream locations from the nozzle exit. And the FFT of the interface location provides the Energy spectrum at those locations. It shows broadband response for unforced cases while clearly identified frequency peaks are observed for forced cases, further supporting the observations from POD modes. It also can be used to examine the evolution of forced jet in crossflow direction. The near nozzle observation point (5 dj) shows the dominance of forcing function, where the energy at other frequencies remains lower. For the downstream location (15 dj), although there is still a dominant peak at the forcing frequency, the energy at other frequencies increases as compared to the near nozzle location.

## Acknowledgements

The authors gratefully acknowledge the support provided by Department of Science and Technology, Government of India through the SERB-SRG grant.

## Acknowledgements

The authors gratefully acknowledge the support provided by Department of Science and Technology, Government of India through the SERB-SRG grant.